\documentclass[journal, twocolumn, 10pt]{IEEEtran}
\usepackage{graphicx}
\usepackage{algorithm}
\usepackage{algorithmicx}
\usepackage{algpseudocode}
\usepackage{cite}
\usepackage{color}
\usepackage{amsmath,amssymb,amsfonts,amsthm,dsfont}
\usepackage{bm}
\usepackage{setspace}
\usepackage{multirow}
\usepackage{mathtools}
\usepackage{epstopdf}

\begin{document}
\title{When Crowdsourcing Meets Mobile Sensing: \\ A Social Network Perspective}
\author{
    \IEEEauthorblockN{Pin-Yu Chen\IEEEauthorrefmark{1}, Shin-Ming Cheng\IEEEauthorrefmark{2}, Pai-Shun Ting\IEEEauthorrefmark{1}, Chia-Wei Lien\IEEEauthorrefmark{3}, Fu-Jen Chu\IEEEauthorrefmark{4}
 } \\
    \IEEEauthorblockA{\IEEEauthorrefmark{1}Department of Electrical Engineering and Computer Science, University of Michigan, Ann Arbor, USA
    \\ \{pinyu, paishun\}@umich.edu} \\
 \IEEEauthorblockA{\IEEEauthorrefmark{2}Department of Computer Science and Information Engineering, National Taiwan University of Science and Technology, Taipei, Taiwan
 	\\ \{smcheng\}@mail.ntust.edu.tw} \\
    \IEEEauthorblockA{\IEEEauthorrefmark{3}Amazon Corporate LLC
    \\ chiaweil@amazon.com} \\
    \IEEEauthorblockA{\IEEEauthorrefmark{4}Institute for Robotics and Intelligent Machines, Georgia Institute of Technology, Atlanta, USA
    \\ fujenchu@gatech.edu} 
}

\maketitle
\setstretch{1.0}
\thispagestyle{empty}
\begin{abstract}
Mobile sensing is an emerging technology that utilizes agent-participatory data for decision making or state estimation, including multimedia applications. This article investigates the structure of mobile sensing schemes and introduces crowdsourcing methods for mobile sensing. Inspired by social network, one can establish trust among participatory agents to leverage the wisdom of crowds for mobile sensing. A prototype of social network inspired mobile multimedia and sensing application is presented for illustrative purpose. Numerical experiments on real-world datasets show improved performance of mobile sensing via crowdsourcing. Challenges for mobile sensing with respect to Internet layers are discussed.
\end{abstract}
%\cite{Albert00}

%\IEEEpeerreviewmaketitle

\section{Introduction}
\label{sec_intro}

\begin{figure}
\centering
\includegraphics[width=3in]{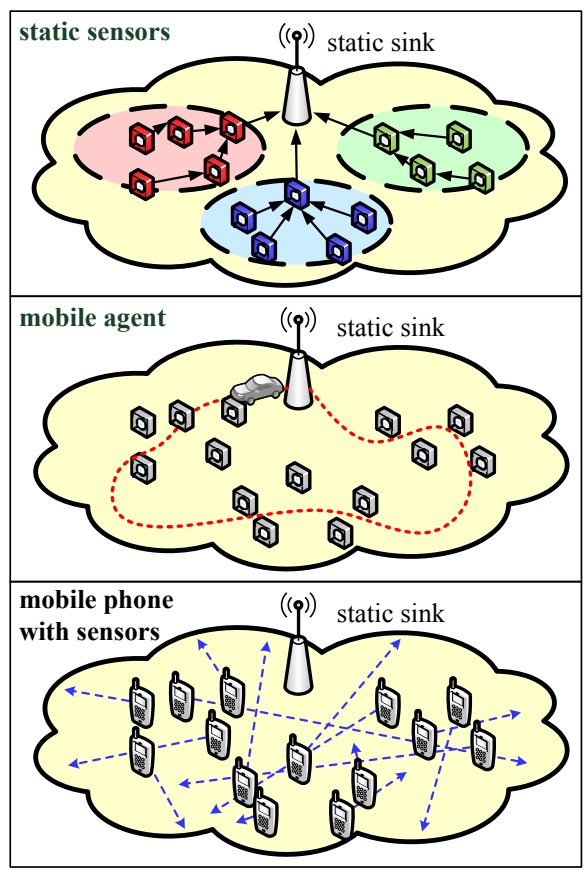}
\caption{Evolution from Wireless Sensor Networks to Mobile Sensing.}
\label{WSNevo}
\end{figure}

Wireless sensor network (WSN) explores the avenues to collect and use information from the physcial world by deploying low-cost tiny sensor nodes on the ground, in the air, under water, on bodies, in vehicles, and inside buildings. With sensing, processing, and communication capabilities, networked sensor nodes cooperatively collect information on entities of interest and WSNs have emerged as a promising technology with numerous and various applications. As shown in Figure~\ref{WSNevo}, sensor nodes locally collect information and then forward the sensed result over a wireless medium to a remote static sink, where it is fused and analyzed in order to determine the global status of the sensed area. In order to successfully gather sufficient information, a static sink could send a mobile agent to collect data from individual sensor nodes by following a trajectory spanning all the nodes (see Figure~\ref{WSNevo}). 
%Consequently, the short distance between sensor nodes and mobile agent dramatically reduces the consuming power.

To accomplish large-scale sensing, WSN evolves not only at the sink side (such as mobile agents), but also at the sensor node side. Mature mobile networks consisted of mobile devices with advanced processing and communication capabilities become a possible sensing infrastructure of WSN. By exploiting the rich set of embedded sensors (such as camera, gyroscope, GPS, accelerator, light sensor and digital compass) on mobile phones as the sensor nodes, a new paradigm of WSN is realized, which is known as \textit{mobile sensing}~\cite{Gaonkar08,Miluzzo08,Mun09,Ganti11,Xiang13,Zhang13,Weppner14,He15,Liu15}. As shown in Figure~\ref{WSNevo}, mobile sensing utilizes crowdsensed information for data analysis and decision making due to penetration of mobile devices as well as human mobility and ubiquity. It relies on the wisdom of crowds \cite{Surowiecki05} to successfully infer the information of interest and accomplish its tasks. The data from the mobile crowds (e.g., users, sensors, robots, e.t.c.) can be either numerical or categorical, depending on applications. Examples of crowdsensed data include numerical environmental measurements such as temperature and air conditions~\cite{Mun09,Xiang13}, personal activities such as daily life patterns and events~\cite{Miluzzo08}, interactions among people such as crowd density~\cite{Weppner14} and common interests~\cite{Zhang13}, categorical recommendations such as ratings for nearby restaurants~\cite{Gaonkar08}, and user experience/quality feedback of wireless mobile multimedia applications~\cite{Gaonkar08}.

It is worth noting that many multimedia applications lie within the scope of mobile sensing, since extracting and analyzing the information sensed or generated from the crowds is one of the core goals for many multimedia applications, in order to attract users' attention.  Better prediction of users' interest leads to longer multimedia stickiness and hence more revenues can be expected. Modern multimedia applications often pull user-centric information from the crowds and offer personalized contents (e.g., next video to watch). Typically, location and social network information are widely used for targeted advertisement and recommendation. Therefore, the major challenge is to efficiently and accurately extract user-centric information from the crowds and identify users of high similarity for improved content delivery.

In recent years, many machine learning tasks and business models have leveraged the wisdom of crowds to acquire crowdsourced data for discriminating unknown objects.  The website \textit{Galaxy Zoo} asks visitors to help classify the shapes of galaxies, and the website \textit{Stardust@home} asks visitors to help detect interstellar dust particles in astronomical images. Business models such as \textit{Amazon Mechanical Turk (MTurk) } and \textit{CrowdFlower} provide crowdsourcing services with low prices. For \textit{MTurk}, a minimum of 0.01 US dollar\footnote{https://requester.mturk.com/pricing} is paid to a labeler/worker when he/she makes a click (i.e., generates a label) for an item. Despite the low costs for acquiring crowdsourced data, one of the major challenges of mobile sensing roots in dealing with noisy and potentially erroneous data \cite{Snow08,Ganti11,He15}. These undesired data can originate from environmental/object uncertainties (e.g., channel noise and difficulty of object discrimination) or user intentions (e.g., fraudulent recommendations and irresponsible user clicks). Consequently, identifying trustworthy data and reliable agents becomes an essential task in mobile sensing  \cite{Ganti11,Xiang13,He15}.

Utilizing the concept of trust in social network, this article proposes trust-based data analysis approaches for crowdsourcing in mobile sensing schemes. In addition to scraping trustworthy data from the bulk, these approaches aim to identifying reliable agents for performance enhancement. A weight of trust is built upon the reliability of each agent for mobile sensing via limited number of training queries. For spectrum sensing, these approaches can be implemented by broadcasting reference signals for reception power calibration. For annotation, these approaches can be implemented by uploading some items with known answers. For multimedia, these approaches can be associated with user behaviors based on the provided contents.

This article summarizes mobile sensing network paradigms and introduces several trust-based crowdsourcing methods for mobile sensing. We also illustrate a prototype application of social network inspired mobile multimedia and sensing scheme. Numerical experiments on real-world datasets show that mobile sensing can benefit from crowdsourcing for improved performance. Potential challenges of social network based mobile sensing with respect to mobile multimedia Internet layers are discussed. This article therefore sheds new light on integration of social network and mobile sensing, and applications therein.

\section{Mobile Sensing Paradigms}
\label{sec_paradigm}

\begin{figure}
\centering
\includegraphics[width=3.3in]{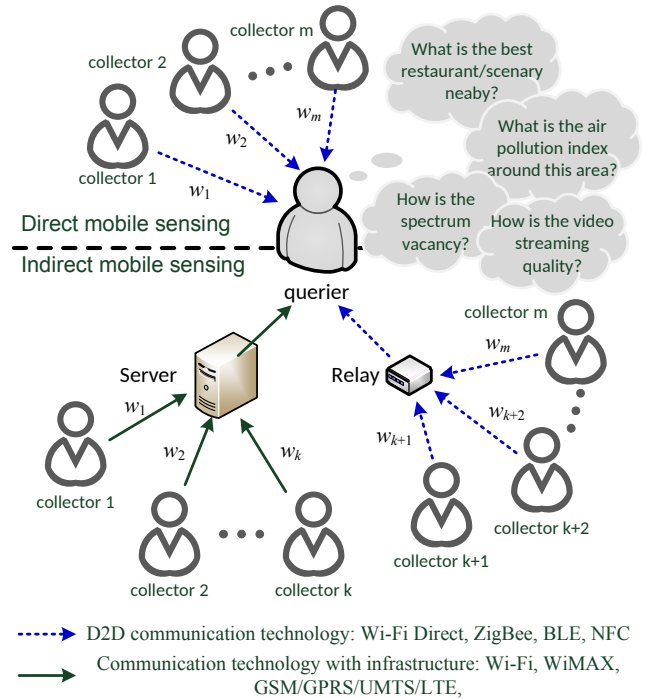}
\caption{Network architecture of mobile sensing.}
\label{fig_arch}
\end{figure}

In mobile sensing, people share and distribute sensed information via physical proximity or social relations over portable sensors. As illustrated in Figure~\ref{fig_arch}, a mobile user plays the roles of both \textit{querier} and \textit{collector}, who respectively requests and provides information. A querier can simultaneously be a collector if he/she also participates in mobile sensing. Network structures of mobile sensing can be classified into two categories, namely the \textit{direct} and \textit{indirect} paradigms, which are described as follows.

\subsection{Direct mobile sensing paradigm} 

A direct mobile sensing paradigm involves direct communication between a querier and crowds (i.e., the collectors). Typically, it is achieved by adopting current device-to-device (D2D) communication technologies, such as Wi-Fi Direct, ZigBee, Bluetooth Low Energy (BLE), or near field communication (NFC). In such a case, the \textit{store-carry-forward} behavior facilitates information delivery in an ad hoc fashion. That is, sensed information could be stored in a sensor node in the absence of immediate connectivity to any other node, and could be relayed to other sensor nodes at encounters. Examples include: 
\begin{itemize}
	\item \textbf{Proximity sensing in mobile social networks (MSN)}: It supports social platforms among physically proximate mobile users. For instance, one can simply scan the environment for discoverable Bluetooth devices to analyze crowd density and crowd flow direction~\cite{Weppner14}. By exploiting P2P communications, one can further make new social interactions with nearby devices. A popular example is sensing ``potential friends with similar interests nearby''. To enjoy such new activities, mobile users have to provide their own interests for profile matching by broadcasting their personal profiles to all nearby users, and then comparing their personal profiles and other users' profiles for friend matching~\cite{Zhang13}. 

	\item \textbf{Cooperative spectrum sensing in cognitive radio networks (CRN)}: Unlicensed secondary users (SUs) sense the surrounding environment and exploit spectrum holes unoccupied by licensed primary users (PUs) for secondary transmission with minimal interference to PUs. To achieve better spectrum management and to enhance radio resource utilization, a querier could exploit observations on local spectrum vacancy from surrounding SUs (i.e., crowds). The empowerment of cooperative spectrum sensing improves the throughput of wireless communications and reduces potential interference among heterogeneous systems.
\end{itemize}

\subsection{Indirect mobile sensing paradigm} 
In this paradigm a querier and crowds are indirectly connected through a communication system in a centralized or a distributed fashion. Typically, access points in WLAN and a base station in cellular network or WiMax are exploited as communication paths in the former case. In the latter case, a querier/collectors could download/upload data from/to nearby relays via localized communication technologies such as wifi-direct, BLE or NFC. Examples include:
\begin{itemize}
	\item \textbf{Environmental measurements}: Collectors provide local measurements  (e.g., temperatures, air pollution indexes, e.t.c.) to a querier via an existing cellular infrastructure for event detection or state estimation. Consequently, the current environment can be understood and be improved~\cite{Mun09,Xiang13}. For example, PEIR project~\cite{Mun09} exploits sensors in mobile phones to build a system that tracks the impact of individual actions on  carbon emissions.

	\item \textbf{Personal activity sharing}: A collector shares his/her daily life patterns, activities (such as sports), health (such as heart rate, blood pressure) with his/her friends using online social networks. For example, by automatically classifying events in people's lives via sensors on mobile phones, CenceMe~\cite{Miluzzo08} enables selective event sharing among friends using Facebook or Twitter.

	\item \textbf{Online recommendation}: Crowds (e.g., data collected from proximal users or users of high similarity) provide recommendations to a user-centric query, such as the best seafood restaurant within 2 miles, or the next video to watch for multimedia applications. For example, Micro-Blog~\cite{Gaonkar08} encourages user to record multimedia blogs manually or automatically (via sensors). Moreover, the blogs from collectors in the same area are integrated to enrich the contents. Consequently, a querier can browse multimedia blogs at a selected region for relevant information.

	\item \textbf{Annotation}: Crowds (e.g., machines, people, e.t.c.) annotate labels, such as scenery labels for a picture or comments and interactions for multimedia contents, for an item inquired by a user. One typical example is the \textit{Amazon Mechanical Turk (MTurk)} service.
\end{itemize}

\begin{figure}[t]
    \centering
    \includegraphics[width=3.5in]{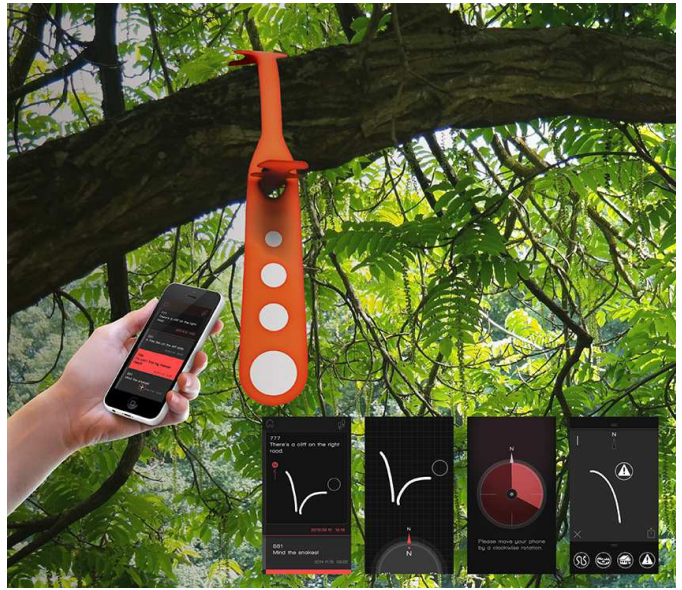}
    \caption{LifeTie mobile sensing system - physical device and its mobile multimedia interface.}
    \label{fig_LifeTie}
\end{figure}

\section{Life Tie: A Prototype Social Network Inspired Mobile Multimedia and Sensing Application}
\label{sec_LifeTie}

For further illustration, in this section we introduce a social network inspired mobile multimedia and sensing application. In Taiwan,  mountainous areas are hikers' paradise. Hikers are used to tie trail marking ribbons on trees for direction guidance. It is a matter of life and death to clearly know one's own location, especially at night time. However, trail marking ribbons have several disadvantages such as misinterpretation, lack of detailed information, and environmental pollution caused by overuse. To ensure hikers' saftey while overcoming the aforementioned drawbacks, we propose a mobile sensing system named ``LifeTie''. Unlike traditional WSN deployed in mountains for wildlife tracking and ecological monitoring, LifeTie acts as an annotation platform where users can exchange their sensing results. The integration of tail marking ribbon and NFC technology replaces the traditional marking method with smartphone APP and achieves the purpose of information exchange, rescue, and search.

The main concept of LifeTie is to exploit NFC tags as the enabler for information exchange among people. To achieve that, NFC tags shall be attached around a mountain, thus creating an infrastructure. Hikers can trigger NFC tags nearby (typical in the range of only a few inches) using their NFC enabled mobile phones. Then the NFC enabled mobile phones can read/write data from/to the NFC tags. With embedded memory, LiftTie could handle a vast amount of data, which explores the possibility of multimedia information and advances toward integration of social network and mobile sensing. The prototype of LifeTie is shown in Figure~\ref{fig_LifeTie}. The features and functionalities of LifeTie are summarized as follows.
\subsection{Features}
\begin{itemize}
\item \textbf{Flexible}: The shape of LifeTie was inspired by zap-straps. It can be easily tied to tree branches. 
\item \textbf{Recognizable}: With the color of fluorescent orange and the addition of reflective stickers, it provides direction guidance even at night time.
\item \textbf{Reliable}: Polypropylene (PP) is used in LifeTie for its bendability and durability. An NFC tag is integrated in the internal parts of a strap to resist extreme weather conditions in mountains. 
\item \textbf{Affordable}: The cost of an NFC tag is low.
\item \textbf{Power-saving}: An NFC tag is not powered by electricity. As a result, no replacement for LifeTie is required, and therefore even lower deployment cost can be achieved.
\end{itemize}
%使用情境

\subsection{Functionalities}
\begin{itemize}
\item \textbf{Navigation and Warning.} Via the corresponding APP, hikers can check out LifeTie's guestbook right after triggering NFC tags. Depending on the current environmental conditions, hikers could leave comments or draw a simple map to make a notification or navigation. Some useful waring icons (such as cliff, snake, wasps, slippery) and guiding icons (such as cave, camp) are provided when drawing the map to enable diverse multimedia contents. The updated surrounding information facilitates the following hikers. 

\item \textbf{Tracking and Rescue.} When a hiker is lost in a mountain and finds LifeTie, he/she can check regular comments to see if there is any shelter nearby. Moreover, he/she can leave urgent comments highlighted by red. If he/she can leave such information on several LifeTies, rescuers can easily identify a rough search area according to the positions of deployed LifeTie and timestamps of the comments. As a result, the rescuing operations become more effective and efficient.
\end{itemize}

\section{Trust-based Crowdsourcing Methods for Mobile Sensing}
\label{sec_method}
\begin{figure*}[t]
	\centering
	\includegraphics[width=5in]{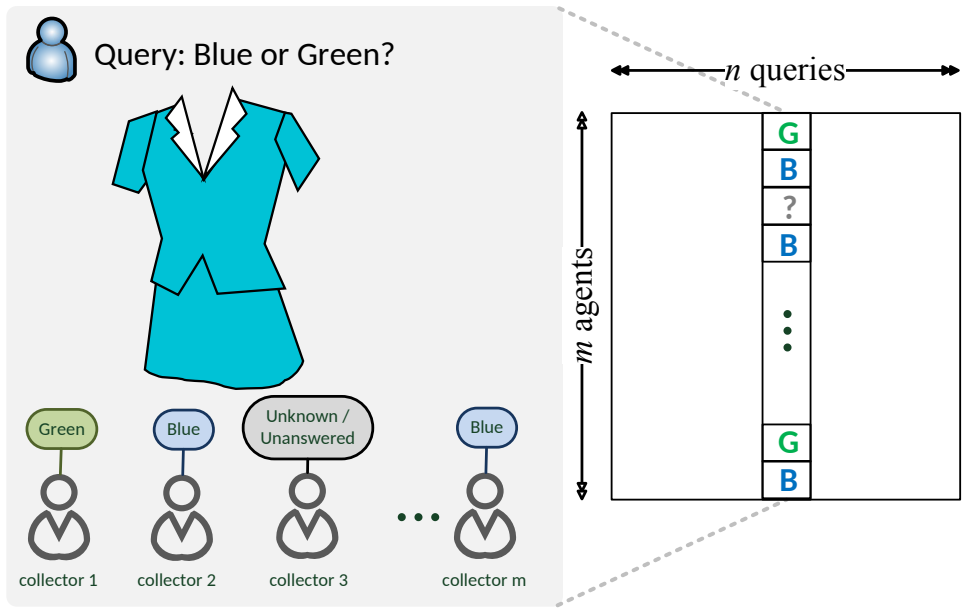}
	\caption{Illustration of crowdsourced data.}
	\label{fig_matrix}
\end{figure*}

This section provides an overview of weight (trust) assignment methodologies on agents for crowdsenced data in mobile sensing. The utility of these methods is investigated in Sec. \ref{sec_num} and the challenges toward practice are addressed in Sec. \ref{sec_cha}.
For crowdsourcing-empowered mobile sensing, a user (or an intermediate system) evaluates a weight of trust $w_i$ for $i$-th agent and fuse information from agents via weighted combination of agents' observations for the user's query. As shown in Figure \ref{fig_matrix}, the collected data from the agents can be viewed as a matrix with rows representing agents and columns representing observations associated with queries. The training queries refer to the queries with known answers and they are used for weight evaluation. The number of agents is denoted by $m$.

The final output for mobile sensing is the combination of each agent's observation multiplied by the associated weight. Here we discuss several crowdsourcing methods involving different weight evaluation approaches. These methods can be classified into two categories, \textit{unsupervised} and \textit{supervised} methods, separated by the need of training queries.

\subsection{Unsupervised crowdsourcing methods}
\begin{itemize}
	\item \textbf{majority votes}: majority votes adopts uniform trust among all agents (i.e., the weight $w_i=1/m$ for all $i$) and selects the observation that most agents agree upon as the final output. This method may lead to poor performance when the majority of agents have incorrect observations or when some observations are maliciously manipulated.
	\item \textbf{probabilistic inference}:	Probabilistic inference method assumes that each observation made by an agent is statistically independent and imposes statistical model to infer the weights from observations. One popular method is the weight evaluation method based on an expectation maximization (EM) algorithm \cite{Raykar10,Xiang13}.
\end{itemize}

\subsection{Supervised crowdsourcing methods}
Supervised crowdsourcing methods aim to find the optimal weight of each agent by solving the  optimization problem as
\begin{align}
\label{eqn_objective}
\text{minimize}_{\mathbf{w}}~&\text{cost(training-queries,~final-output)} 
+\lambda \cdot R(\mathbf{w}), \nonumber
\end{align}
where $\mathbf{w}=[w_1,w_2,\ldots,w_m]$ is the vector of weights, $R(\mathbf{w})$ is a regularization function for $\mathbf{w}$, and $\lambda \geq 0$ is the regularization parameter for $R(\mathbf{w})$. Here we introduce several supervised crowdsourcing methods. 
\begin{itemize}
	\item \textbf{weighted averaging} weighted averaging is a heuristic weight evaluation approach which assigns weight that is proportional to the accuracy of each agent in the training queries.
	Let $q_i$ be the fraction of correct queries responded by agent $i$. The weight of agent $i$ is the normalized accuracy 
	$w_i=q_i/{\sum_{i=1}^m q_i}$.

	\item \textbf{exponential weighted algorithm}:	
	exponential weighted algorithm adopts exponential cost function and zero regularization parameter ($\lambda=0$) and sequentially adjusts the weight of each agent from the training queries. Interested readers can refer to \cite{Hastie01} and the references therein for more details.
	
    \item \textbf{support vector machine}: support vector machine adopts the Hinge loss function as its cost function and assumes the regularization function $R(\mathbf{w})=\sum_{i=1}^m w_i^2$ and positive regularization parameter (i.e., $\lambda >0$). Support vector machine aims to find a separating hyperplane that best discriminates the training queries in the data sample space, and the weight of each agent can be determined by the resulting separating hyperplane. Interested readers can refer to \cite{Hastie01} and the references therein for more details.
	
	\item \textbf{professional search}: Inspired by social networks where problems are often resolved by professionals, professional search aims to assign weights to only a few agents that have outstanding accuracy in the training queries \cite{CPY15crowd}. Professional search adopts the Hinge loss function as its cost function and assumes the regularization function $R(\mathbf{w})=\sum_{i=1}^m |w_i|$. This regularization function is known as a surrogate function that promotes sparsity in $\mathbf{w}$
	(i.e., most of the weights are zero), and hence the professionals hidden in the crowds can be selected for mobile sensing.
\end{itemize}

\section{Numerical Experiments}
\label{sec_num}
In this section we use two crowd generated datasets to investigate the performance of the crowdsourcing methods in Sec. \ref{sec_method}. For the first dataset, each agent only participates in some fraction of queries and hence it resembles the dynamic participatory nature in mobile sesing. For the second dataset almost every agent responds to each query but none of the agents have correct answers to all queries, which resembles the imperfect sensing capability in mobile sensing.
In both scenarios, crowdsourcing methods can improve the query classification performance by identifying trustworthy agents.

For crowdsourcing methods involving a regularization function $R$, we use leave-one-out-cross-validation (LOOCV) approach \cite{Hastie01} to determine the optimal regularization parameter $\lambda$, by 
swiping $\lambda$ from $0$ to $200$ to select the optimal value that leads to minimum training error. One-to-all classification approach is used for multiple (more than two) categorical datasets (e.g., the exam dataset).

\subsection{Text relevance judgment}
The text relevance judgment dataset is provided by the text retrieval conference (TREC) crowdsourcing track in $2011$\footnote{https://sites.google.com/site/treccrowd/2011}, where $689$ agents (participants) are asked to judge the relevance of paragraphs excerpted from a subset of articles with given topics. Each agent then generates an observation, either ``relevant'' or ``irrelevant'', for an article. It is worth mentioning that 
this dataset is sparse in the sense that in average each agent only read roughly $26$ out of $394$ articles. For supervised crowdsourcing methods we use roughly $10\%$ ($40$ training queries) of articles to evaluate each agent's weight. The rest of data samples are used to test the accuracy of crowdsourcing algorithms and the results are summaried in Table~\ref{table_TREC}. It is observed that supervised methods can achieve higher accuracy than unsupervised methods via training queries. Also note that support vector machine and professional search outperform other methods since their main objective is to assign more weights on the trustworthy agents/data samples possessing eminent discriminant capability.

\begin{table*}[t]
	\label{table_TREC}
	\caption{The TREC 2011 dataset. Supervised methods attain higher accuracy than unsupervised methods via acquiring a few training queries for weight (trust) assignment.}
	\centering
\begin{tabular}{c|cc|cccc}
	\multicolumn{1}{l|}{}                                     & \multicolumn{2}{c|}{\begin{tabular}[c]{@{}c@{}}unsupervised\\ ~\end{tabular}}                                                   & \multicolumn{4}{c}{\begin{tabular}[c]{@{}c@{}}supervised\\ ~\end{tabular}}                                                                                                                                                                                                      \\ \hline
	Methods                                                   & \begin{tabular}[c]{@{}c@{}}majority \\ votes\end{tabular} & \begin{tabular}[c]{@{}c@{}}expectation \\ maximization\end{tabular} & \begin{tabular}[c]{@{}c@{}}weighted \\ averaging\end{tabular} & \begin{tabular}[c]{@{}c@{}}exponential\\ weighted\\ algorithm\end{tabular} & \begin{tabular}[c]{@{}c@{}}support\\ vector\\ machine\end{tabular} & \begin{tabular}[c]{@{}c@{}}professional\\ search\end{tabular} \\ \hline
	\begin{tabular}[c]{@{}c@{}}~\\ Accuracy (\%)\end{tabular} & \begin{tabular}[c]{@{}c@{}}~\\ 79.38\end{tabular}         & \begin{tabular}[c]{@{}c@{}}~\\ 78.81\end{tabular}                   & \begin{tabular}[c]{@{}c@{}}~\\ 83.05\end{tabular}             & \begin{tabular}[c]{@{}c@{}}~\\ 80.51\end{tabular}                          & \begin{tabular}[c]{@{}c@{}}~\\ 83.33\end{tabular}                  & \begin{tabular}[c]{@{}c@{}}~~\\ 84.46\end{tabular}           
\end{tabular}
\end{table*}

\begin{table*}[t]
	\label{table_exam}
	\caption{The science exam dataset. Despite the fact that no students answer all questions correctly,
		professional search can still achieve more than twice of accuracy than random guess (25\% accuracy). } 
	\centering
	\begin{tabular}{c|cc|cccc}
		\multicolumn{1}{l|}{}                                     & \multicolumn{2}{c|}{\begin{tabular}[c]{@{}c@{}}unsupervised\\ ~\end{tabular}}                                                   & \multicolumn{4}{c}{\begin{tabular}[c]{@{}c@{}}supervised\\ ~\end{tabular}}                                                                                                                                                                                                      \\ \hline
		Methods                                                   & \begin{tabular}[c]{@{}c@{}}majority \\ votes\end{tabular} & \begin{tabular}[c]{@{}c@{}}expectation \\ maximization\end{tabular} & \begin{tabular}[c]{@{}c@{}}weighted \\ averaging\end{tabular} & \begin{tabular}[c]{@{}c@{}}exponential\\ weighted\\ algorithm\end{tabular} & \begin{tabular}[c]{@{}c@{}}support\\ vector\\ machine\end{tabular} & \begin{tabular}[c]{@{}c@{}}professional\\ search\end{tabular} \\ \hline
		\begin{tabular}[c]{@{}c@{}}~\\ Accuracy (\%)\end{tabular} & \begin{tabular}[c]{@{}c@{}}~\\ 46.67\end{tabular}         & \begin{tabular}[c]{@{}c@{}}~\\ 50\end{tabular}                      & \begin{tabular}[c]{@{}c@{}}~\\ 46.67\end{tabular}             & \begin{tabular}[c]{@{}c@{}}~\\ 26.67\end{tabular}                          & \begin{tabular}[c]{@{}c@{}}~\\ 50\end{tabular}                     & \begin{tabular}[c]{@{}c@{}}~~\\ 53.33\end{tabular}           
	\end{tabular}
\end{table*}

\subsection{Science exam dataset}

The science exam dataset is collected by the authors and it contains $40$ questions. Each question has four choices and the correct answer is one of these four choices. There are $183$ agents (students) taking the exam and producing observations (their answers). Unlike the TREC dataset, this exam dataset is dense in the sense that almost every student has provided an answer for each question. We use $10$ questions as training queries and use the rest to test the accuracy. The results are summarized in Table~\ref{table_exam}. The baseline accuracy (random guess) is $25\%$. None of the methods can achieve accuracy as high as in the TREC dataset due to the following facts. 
\begin{enumerate}
\item The observations of TREC dataset only have two categories, whereas the observations of the exam dataset have four categories, which renders the latter more difficult to be discriminated.  
\item The exam is challenging since majority votes method leads to low accuracy and there is not a student who answers all questions correctly. The best student only has $70\%$ accuracy. 
\end{enumerate}
Nonetheless, crowdsourcing methods such as support vector machine and professional search can still attain relatively good accuracy by identifying reliable agents.

\section{Ongoing Challenges: Aspects from Mobile Multimedia Internet Layers}
\label{sec_cha}
In this section we discuss some ongoing challenges toward integration of social network and mobile sensing, particularly on the aspects of mobile multimedia Internet. Issues corresponding to each layer of mobile multimedia Internet are specified as follows.

\subsection{Application Layer}
\begin{itemize}
\item \textbf{Conflict between privacy and trustworthiness.} Although such an integration of social network and mobile sensing is exciting and promising, the collection and sharing of personal information related to human activity introduces the concern of privacy, where the participants are reluctant to reveal any sensitive personal information (such as time, location, pictures, sound, acceleration, and biometric data). As a result, it is crucial to design an approach to collecting sufficient information from participants without violating their privacy~\cite{He15}. Specifically, {\it authentication} shall be supported to identify legal mobile users and adversaries. Moreover, {\it anonymity} shall be preserved to hide sensitive information by using technology such as k-anonymous or pseudonyms. These avenues prevent adversary from traversing relationship between users' contributions and identities.

In particular, trust-based crowdsourcing methods aim to preserve trustworthiness in harsh environment where malicious participants may deliberately feedback fraudulent data. Obviously, to counteract this effect we need to observe contributions made by each user for a period of time and hence to evaluate his/her trustworthiness. However, it may conflict with the privacy consideration where actual attribute values of a specific user are obscured and the links between multiple contributions from the same user are broken. How to acquire linkability across multiple contributions from the same user while preserving privacy is a challenging issue~\cite{Xiang13}.

\item \textbf{Data integrity on complicated multimedia content.} By manually recording via users or automatically collecting via sensors, a huge amount of information can be retrieved in mobile sensing. The multimedia contents generated from the retrieved  materials via application like Micro-Blog~\cite{Gaonkar08}  contain  abundant yet complicated information, which burdens data integrity. Unlike the example raised in Figure~\ref{fig_matrix} where we just need to identify the color of a cloth, multiple parts in one clip or video might lead to distinct conclusion that is highly dependent upon a viewer's ideology. Extra meta information shall also be included to increase data integrity like user reviews (such as IDMB, Youtube) or user preference (such as Netflix).

\item \textbf{Incentives of participation.} Mobile sensing requires participants to spend their time, attention, and mobile phone's battery power for contributing data. Obviously, the amount of collected information and the number of participants are proportional to the degree of voluntary. When the amount of collected information is insufficient, the sensing results might not be precise. Thus, mobile sensing needs incentive and mechanism design to encourage people to participate~\cite{Yang12}.
\end{itemize}

\subsection{Network Layer}
\begin{itemize}
\item \textbf{Data retrieval in distributed environment.} In distributed scenarios (such as proximity sensing, spectrum sensing, or annotation), a querier can only retrieve information from localized collectors, which might lead to biased inference results. To overcome this issue, current researches propose to enable \textit{information relay} for each individual. Leveraging human mobility and the store-and-forward features, the amount of data collected from crowds grows substantially via information exchange, thereby improving the accuracy of estimation. In addition, in indirect mobile sensing paradigm involving data retrieval and analysis from distributed systems (e.g., data storage servers), distributed computation is known to be one of the \textit{big data} challenges.

\item \textbf{The limitations of D2D communications.} In the direct sensing paradigm, a querier could communicate with collectors via D2D communications like Wi-Fi Direct, ZigBee, BLE, and NFC. However, current D2D communication technologies typically require manual mutual-authentication when making a connection, which is unfavorable for automatic data collection and device connectivity. Moreover, mobile sensing applications should be capable of integrating the features of different D2D communication technologies (such as transmission ranges, transmission rates, and power consumptions) for ubiquitous multimedia content delivery.
\end{itemize}

\subsection{Link Layer}
\begin{itemize}
\item \textbf{Unreliable link due to mobility and interference.} The mobile nature of users and agents may hinder the performance of mobile sensing due to change in location, environment, and participatory agents. The highly dynamic positions of agents incur ever-changing received interference, thereby affecting the link reliability. To accommodate this effect, a crowdsourcing aided mobile sensing method should possess adaptivity and robustness in such a dynamical situation in order to identity inadequate agent participation and obsolete data collection. 
\end{itemize}

\subsection{Physical Layer}
\begin{itemize}
\item \textbf{Tradeoff between power consumption and sensing accuracy.} Apparently, the higher sensing accuracy is attained, the more energy is consumed in mobile devices with limited power due to the increase in data acquisition frequency, which might violate the design rationale of sensing paradigms. Consequently, power consumption fairness and energy-efficient scheduling for participatory devices should be jointly considered and new throughput measures should be studied to balance the tradeoff between power consumption and sensing accuracy.
\end{itemize}

\section{Conclusion}
This article propose to incorporate crowdsourcing methods for mobile sensing and introduces several crowdsourcing methods for evaluating the weight of trust among agents. The direct and indirect mobile sensing network paradigms are discussed. A prototype social network inspired mobile multimedia and sensing application is illustrated toward integration of social network and mobile sensing. Numerical experiments on real-world datasets show that mobile sensing can benefit from crowdsourcing methods for performance improvements. Ongoing challenges of integration of social network and mobile sensing are addressed from the aspects of mobile multimedia Internet layers. This article therefore paves new avenues to various mobile applications and future mobile technology development.

\bibliographystyle{IEEEtran}
\bibliography{IEEEabrv,crowdsourcing}

\end{document}